\documentclass[aps,pra,twocolumn,groupedaddress,showpacs,showkeys]{revtex4}

\usepackage{graphicx}
\begin{document}

\newcommand{\be}{\begin{equation}}
\newcommand{\ee}{\end{equation}}

\title{ New classical  brackets for dissipative systems.}

\author{ G. Bimonte, G. Esposito, G. Marmo and C. Stornaiolo}
\email[Bimonte@na.infn.it, Giampiero.Esposito@na.infn.it,
Gimarmo@na.infn.it,  Cosmo@na.infn.it]
\affiliation{Dipartimento di Scienze Fisiche, Universit\`{a} di
Napoli Federico II, Complesso Universitario MSA, Via Cintia
I-80126 Napoli, Italy; INFN, Sezione di Napoli, Napoli, ITALY.}

\date{\today}

\begin{abstract}
A set of brackets for  classical dissipative systems, subject to
external random forces, are derived. The method is inspired to the
old procedure found by Peierls, for deriving the canonical
brackets of conservative systems, starting from an action
principle. It is found that an adaptation of Peierls' method is
applicable also to dissipative systems, when the friction term can
be described by a linear functional of the coordinates, as is the
case in the classical Langevin equation, with an arbitrary memory
function.  The general expression for the brackets satisfied by
the coordinates, as well as by the external random forces, at
different times, is determined, and it turns out that they all
satisfy the Jacobi identity. Upon quantization, these classical
brackets  are found to coincide with the commutation rules for the
quantum Langevin equation, that  have been obtained in the past,
by appealing to microscopic conservative quantum models for the
friction mechanism.

\end{abstract}

\pacs{05.30.-d, 03.65.Db}
\keywords{brackets, dissipation, Langevin, quantization}

\maketitle
\section{Introduction}

The study of quantum dissipative systems, has attracted, in the
last decades, a lot of interest, in view of its broad spectrum of
applications, ranging from quantum optics through statistical
mechanics, etc.  The standard approach to deal with  quantum
dissipation, is based on the idea that the physical origin of
dissipation is the interaction of the system with a heat bath,
consisting of a large number of degrees of freedom. One considers
then some microscopic, conservative model for the heat bath (and
its interaction with the system), and tries  to recover the
macroscopic quantum behavior of the dissipative system alone, by
eliminating from the description the degrees of freedom describing
the bath. In Ref.(\cite{ford}), it is shown, indeed, that the most
general quantum Langevin Equation,  which is  one of the most
popular models for dissipation,   can be obtained from a simple
microscopic model, where the heat bath is described by a set of
independent oscillators, linearly coupled to the system of
interest.

One may wonder whether it is possible to find a quantization
method for dissipative systems, which is based   {\it on the
macroscopic description of dissipation only}, and makes therefore
no use of microscopic models.  As is well known, quantization of
dissipative systems is by no means straightforward, because in
general they admit neither a Lagrangian nor a Hamiltonian
formulation. Moreover, even in those special instances where such
a formulation can be given, the application  of the conventional
canonical quantization methods leads to physically incorrect
results \cite{senit}.  In this Letter, we show that new classical
brackets can be consistently built for dissipative systems, by
generalizing the covariant definition of Poisson Brackets for
Lagrangian systems, discovered long ago by Peierls \cite{peier}
(See also Refs.\cite{dewitt,maro,duts,bimonte}). Our bracket is
defined on the infinite-dimensional functional space consisting of
all possible classical trajectories, that are accessible to the
system under the influence of the random force. It turns out that,
when dissipation is present, the random external force has a
non-vanishing bracket with the system coordinates, which implies
that it cannot be consistently taken to be zero. This seems to be
in agreement with the fluctuation-dissipation theorem, which
requires fluctuating forces, in the presence of dissipation.

By the correspondence principle, our classical brackets can be
eventually quantized, upon  substituting them by ($1/(i  \hbar)$
times) commutators. In this way, we recover the same expressions
for the commutators between the system coordinates and the random
forces, which were derived from the independent oscillator
microscopic model of Ref.\cite{ford}.

In  what follows, we make no attempt at mathematical rigor, and
the presentation is totally heuristic. We hope to clarify
elsewhere the delicate issues involved in the consideration of
Poisson structures in infinite-dimensional functional spaces.

\section{The classical brackets}

We consider a mechanical system, with coordinates
$(q^1,\cdots,q^n)$, described by an action  functional  $S=\int dt
\;{\cal L}(q^i,\dot q ^i, t)$, where dot denotes a time
derivative. We assume that the Lagrangian is a  polynomial of
second degree in the velocities $\dot q ^i$, and that its Hessian
$ {\partial ^2 {\cal L}}/{\partial \dot q ^i
\partial \dot  q^ j}$ is a constant, non-degenerate matrix
$M_{ij}$. We imagine that the system is in contact with a heat
bath, and we assume that the influence of the heat bath can be
described, effectively, by a mean force, characterized by a
bounded memory function $\mu_{ij}(t-t')$, and a random force
$F_i(t)$. The time  evolution of the system is then described by
the following equation of Langevin type: \be -\frac{\delta
S}{\delta q^i(t)} +
  \int_{-\infty}^t dt'\,
\mu_{ij}(t-t')\;\dot q ^j (t')=F_i(t)\;.\label{lang} \ee Here,
$\delta S/\delta q^i(t)$ denotes the functional derivative of the
action $S$: \be \frac{\delta S}{\delta q^i(t)} \equiv
-\frac{d}{dt}\left(\frac{\partial {\cal L}}{\partial
\dot{q}^i}\right)+\frac{\partial {\cal L}}{\partial q^i} \;.\ee
The original form of the Langevin Equation results from the
singular limit, where $\mu_{ij}(t-t')$ approaches
$\gamma_{ij}\delta (t-t')$.

Mimicking the procedure found by Peierls, to compute the Poisson
Brackets of a conservative Lagrangian system \cite{peier}, one can
consider the effect, on the system evolution, of a small
disturbance, produced by an infinitesimal change $\bar{\delta} S$
in the form of the action. We consider changes of the form
$\bar{\delta} S= \epsilon A$, where $\epsilon$ is an infinitesimal
constant and $A$ is a local functional of the trajectory $q^j(t)$,
taken from a finite time interval. The small disturbance causes an
infinitesimal shift $\delta_A q^j(t)$ in the trajectory $q^j(t)$,
and it is easy to see that, to first order in $\epsilon$,
$\delta_A q^j$ satisfies the following linear integro-differential
equation:
\begin{widetext} \be (L\; \delta_A q)_i(t)\equiv -\int dt' \; \frac{\delta^2 S}{\delta q^i(t)
\delta q^j(t')} \; \delta_A q^j(t')
+\int_{-\infty}^t dt'\, \mu_{ij}(t-t')\; \delta_A \dot q ^j
(t')=\epsilon \frac{\delta A}{\delta q^i(t)}\;, \label{dist}\ee
\end{widetext}
where it is understood that all functional derivatives are
evaluated along the undisturbed  trajectory. When writing the
above Equation, we have also assumed that the random force does
not undergo any variation, to first order in $\epsilon$. We point
out that, by virtue of our assumptions on the Lagrangian, the
coefficients of Eq.(\ref{dist}) depend only on the coordinates
$q^i(t)$ and the velocities $\dot q ^i(t)$ of the undisturbed
trajectory, while they are independent of the accelerations. This
is  reassuring, because,   by virtue of the   random external
force, the classical trajectories possess, in general, smooth
velocities, while the acceleration does not exist, in the ordinary
sense of time-derivatives of the velocity \cite{nelson}.

Since the disturbance $A$ is localized in a finite time interval,
it makes sense to consider the solution $\delta ^- _A q^j(t)$ of
Eq.(\ref{dist}) satisfying {\it retarded} boundary conditions: \be
 \delta ^- _A q^j(t)=0 \;\;\;\;{\rm at \;early \;times}\;.\ee The
non-degeneracy condition for the Hessian $M_{ij}$ of the
Lagrangian, ensures that $\delta ^- _A q^j(t)$ exists and is
unique. We consider also the {\it advanced} solution $\delta ^+ _A
q^j(t)$: \be \delta ^+ _A q^j(t)=0\;\;\;\;\;{\rm at\; late\;
times}\ee of the {\it adjoint} equation of Eq.(\ref{dist}):
\begin{widetext}\be  (L^T\; \delta_A^+ q)_i(t)=-\int dt' \;  \frac{\delta^2
S}{\delta q^j(t') \delta q^i(t)} \delta^+_A q^j(t')\;
- \int_{t}^{\infty} dt'\,\mu_{ji}(t'-t)\,\delta^+_A \dot q ^j (t')
\; =\epsilon \frac{\delta A}{\delta q^i(t)}\;, \label{distad}\ee
\end{widetext} where the superscript $T$ stands for transpose
(the transpose coincides with the adjoint, because we are in the
real field). If $B$ is another functional of the trajectory, with
support in a finite time interval, we define the bracket $\{A,B\}$
as the following expression, involving the quantities
$\delta^{\pm}_A q^j(t)$: \be \{A,B\}:= \frac{1}{\epsilon}\int dt
\frac{\delta B}{\delta q^i(t)} (\delta^+_A q^i(t)-\delta^-_A
q^i(t))\;.\label{brack} \ee It is immediate to verify  that the
bracket is bilinear and satisfies the Leibniz rule:
  \be\{AB,C\}=\{A,C\}\,B+A\{B,C\}\;,\ee \be
\{A,BC\}=\{A,B\}\,C+B\{A,C\}\;.\ee   To verify that the bracket
Eq.(\ref{brack}) is also antisymmetric and that it satisfies the
Jacobi identity, it is useful to reexpress it in terms of the
Green functions $G^{\pm ij}(t,t';q)$, defined so that: \be
\delta^{\pm}_A q^i(t)= \epsilon \int dt' \, G^{\pm ij}(t,t';q)
\frac{\delta A}{\delta q^j(t')}\;.\ee The Green functions $G^{\pm
ij}(t,t')$ satisfy the following boundary conditions: \be
G^{-ij}(t,t';q)=0 \;\;,\;\;\;{\rm for} \;\;t \le
t'\;,\;\;,\label{bm}\ee \be G^{+ij}(t,t';q)=0 \;\;,\;\;\;\;{\rm
for} \;\;t \ge t'\;,\label{bp}\ee
\be \lim_{t \rightarrow t'^{\mp}} \frac{\partial G^{\pm
ij}}{\partial t}(t,t';q) =\mp (M^{-1})^{ij}\;.\label{bpm}\ee We
define now the {\rm commutator function}
$\widetilde{G}^{ij}(t,t';q)$: \be \widetilde{G}^{ij}(t,t';q):=G^{+
ij}(t,t';q)-G^{-ij}(t,t';q)\;.\ee Note that, by virtue of the
boundary conditions satisfied by the retarded and the advanced
Green functions, $\widetilde{G}^{ij}(t,t')$ and $\partial _t
\widetilde{G}^{ij}(t,t')$ are continuous, in the coincidence time
limit, $t \rightarrow t'$. By using $\widetilde{G}^{ij}(t,t')$, we
can rewrite the bracket Eq.(\ref{brack}) as: \be \{A,B\}=
 \int dt  \int dt'\, \frac{\delta B}{\delta
q^i(t)}\,\widetilde{G}^{ij}(t,t';q)\, \frac{\delta A}{\delta
q^j(t')} \;.\label{brack2} \ee The antisymmetry of the bracket
Eq.(\ref{brack}) follows from the fact that the commutator
function $\widetilde{G}^{ij}$ is antisymmetric, as a consequence
of the following reciprocity relation, satisfied by the advanced
and retarded Green functions: \be G^{+ ij}(t,t';q)=G^{-
ji}(t',t;q)\;.\label{recip}\ee Before turning to the proof of
Eq.(\ref{recip}), it is useful to  introduce the condensed index
notation devised by DeWitt \cite{dewitt}. With this notation, the
trajectory $q^i(t)$ is just denoted as $q^i$, with the single
latin index $i$ playing the r\^ole of both the discrete index, and
the time variable. Consequently, repeated condensed indices mean a
summation on the discrete indices as well as a time integration.
For example, Eq.(\ref{dist}), with the condensed notation, is
written as: \be L_{ij}\,\delta_A
q^j\equiv(-S,_{ij}+\kappa_{ij})\,\delta_A q^j=\epsilon
A,_i\;,\label{cond}\ee where commas denote functional
differentiation, and $\kappa_{ij}\, \delta_A q^j$ is a symbolic
notation for the integral linear operator, depending on the memory
function,
 in Eq.(\ref{dist}).  To prove the reciprocity relation
Eq.(\ref{recip}), we point out that the Green functions satisfy by
definition the following Equations: \be L_{ij}\,
G^{-jk}=\delta_i^k\;,\;\;\;\;\;\;(L^T)_{ij}\,G^{+
jk}=\delta_i^k\;.\label{eqgr}\ee Upon multiplying the second of
the above Equations by $G^{-il}$, we obtain: \be
G^{-il}(L^T)_{ij}\,G^{+
jk}=G^{-il}\delta_i^k=G^{-kl}\;.\label{reco}\ee However, upon
using the first of Eq.(\ref{eqgr}), we can rewrite the l.h.s. of
the above Equation as: \be G^{-il}(L^T)_{ij}\,G^{+
jk}=G^{-il}(L)_{ji}\,G^{+ jk}=\delta_j^l \,G^{+ jk}=G^{+
lk}\;.\label{rect}\ee  Upon comparing the r.h.s. of
Eq.(\ref{reco}) and  Eq.(\ref{rect}), the reciprocity relation
Eq.(\ref{recip}) follows. We can verify now the Jacobi identity.
Direct evaluation of the quantity $\{\{A,B\},C\}+
\{\{C,A\},B\}+\{\{B,C\},A\}$, using Eq.(\ref{brack2}) shows that:
\be \{\{A,B\},C\}+{\rm c.p.}=A,_i B,_j C,_k {\cal T}^{ijk}\;,\ee
where c.p. stands for cyclic permutations of the functionals $A,
B, C$.  The terms involving second order functional derivatives of
$A$, $B$ and $C$ cancel exactly,  by virtue of the antisymmetry of
$\widetilde{G}^{ij}$. In the above expression, ${\cal T}^{ijk}$
denotes the following quantity, constructed out of functional
derivatives of the commutator function: \be {\cal
T}^{ijk}=\widetilde{G}^{il}\widetilde{G}^{jk}\!,_l+\widetilde{G}^{jl}\widetilde{G}^{ki}\!,_l+
\widetilde{G}^{kl}\widetilde{G}^{ij}\!,_l\;.\label{tau}\ee By
using the reciprocity relation, the quantity ${\cal T}^{ijk}$ can
be written solely in terms of the retarded Green function
$G^{-ij}$, and its functional derivatives. On the other hand, the
functional derivatives ${G}^{-jk}\!,_l$ can be computed by
functionally differentiating the first of Eqs.(\ref{eqgr}): \be
L_{ij},_l G^{-jk}+L_{ij}\, G^{-jk}\!,_l=0\;.\ee Multiplication by
$G^{+mi}$ then gives: \be G^{-mk}\!,_l=-G^{+mi} L_{ij},_l
G^{-jk}\;=-G^{-im} L_{ij},_l G^{-jk},\ee where in the last passage
use has been made again of the reciprocity relation. By using this
expression into Eq.(\ref{tau}), it is possible to verify that: \be
{\cal T}^{ijk}=(G^{-li}G^{-mj}G^{-nk}+{\rm c.p.}
)(L_{mn},_l-L_{nm},_l)\;,\ee where c.p. stands for cyclic
permutations of the indices $ijk$. It is easy to check now that
${\cal T}^{ijk}$ vanishes. Indeed, in view of Eq.(\ref{cond}), we
see that the quantity between the brackets of the r.h.s. is equal
to: \be S,_{mnl}-S,_{nml}+ \kappa_{mn},_l-\kappa_{nm},_l\;.\ee The
terms involving third order functional derivatives of the action
functional cancel each other, because functional derivatives
commute with each other. On the other hand, the quantities
$\kappa_{ij}$ are independent of the trajectories $q^j$, and hence
their functional derivatives vanish identically. It follows then
that ${\cal T}^{ijk}$ vanishes, and thus the Jacobi identity
holds.\\ Thus we have shown that it is possible to define a
bracket on the space of all trajectories. We can evaluate now the
brackets satisfied by the random force $F_i(t)$. To do this, we
can use the expression for $F_i(t)$, provided by the Langevin
Equation, Eq.(\ref{lang}). In this way, we find:
\begin{widetext}$$ \{F_i, q^k \}= \{-S,_i+\kappa_{ij}\,q^j,q^k
\}=L_{ij}\{q^j,q^k\}=L_{ij}(G^{+jk}-G^{-jk})=(L-L^T)_{ij}\,G^{+jk}+
(L^T)_{ij}G^{+jk}-L_{ij} G^{-jk}=$$ \be
=(S,_{ij}-S,_{ji}+\kappa_{ij}-\kappa_{ji})\,G^{-kj}=(\kappa-\kappa^T)_{ij}\,G^{-kj}\;,\ee
\end{widetext}
and
\begin{widetext}
$$\{F_i, F_j\}=\{-S,_i+\kappa_{ik}\,q^k,
-S,_j+\kappa_{jl}\,q^l
\}=L_{ik}L_{jl}\{q^k,q^l\}=L_{ik}L_{jl}({G}^{+kl}-G^{-kl})=
$$
\be
=L_{ik}L_{jl}({G}^{-lk}-G^{-kl})=L_{ij}-L_{ji}=S,_{ij}-S,_{ji}+
\kappa_{ij}-\kappa_{ji}=(\kappa-\kappa^T)_{ij}\;.\ee
\end{widetext}
It is useful to write the above bracket in plain form: \be
\{F_i(t), F_j(t')\}=\frac{d \mu_{ij}}{dt}(t-t')+\frac{d
\mu_{ji}}{d t}(t'-t)\;.\ee We see from these Equations that, when
friction is present, the external forces have non-vanishing
brackets, which  implies that they cannot be set to zero.

Using Eq.(\ref{brack2}), it is possible to verify that the
equal-time brackets for the coordinates $q^j(t)$ and the momenta
$p_i(t) \equiv \partial{\cal L}/{\partial \dot q ^i(t)}$ have the
familiar canonical form: \be \{q^i(t),q^j(t)\}=0\;,\label{qq}\ee
\be \{q^i(t), p_j(t)\}=\delta_j^i\;,\label{qp}\ee \be
\{p_i(t),p_j(t)\}=0\;.\label{pp}\ee The verification is similar to
the conservative case \cite{dewitt}, because the memory function
contributes to $\widetilde{G}^{ij}(t,t')$ only to order
$(t-t')^3$. This can be seen by inserting the expansions of
${G}^{\pm ij}(t,t')$ in powers of $(t-t')$ into Eqs.(\ref{eqgr}),
and exploiting the boundedness of the memory function.

\section{Conclusions}

We have constructed a set of brackets for a classical dissipative
system, described by a Langevin Equation, with an arbitrary memory
function. The brackets satisfy the usual properties enjoyed by
Poisson Brackets of Hamiltonian systems. It is worth pointing out
the essential r\^ole played, in our treatment,  by external random
forces. When dissipation occurs, they have non-vanishing brackets
with the system coordinates, and thus cannot be consistently set
to zero. As a result, our brackets are a priori defined on the
infinite-dimensional functional space of all possible
trajectories, accessible to the system under the action of
arbitrary external forces. However, in the absence of friction,
when the dynamics is conservative, the brackets can be restricted
onto the finite-dimensional classical phase-space, spanned by the
solutions of the classical Equations of motion, with no external
forces. In this case, our construction reproduces  Peierls'
covariant definition of the Poisson Brackets, for dynamical
systems admitting an action principle. In the context of
conservative systems, the possibility of extending the brackets
from the phase space to the space of all trajectories, was
considered some-time ago \cite{maro}, and our brackets coincide
with those of Ref.\cite{maro}, in the absence of friction.

Quantization can be carried out according to the traditional
procedure, by replacing the classical brackets with commutators.
The resulting commutation rules coincide with those that are
obtained in standard treatments of quantum dissipation, by making
recourse to a microscopic model for the heat bath, after
elimination of the bath degrees of freedom (see for example
Ref.\cite{ford} and references therein).

G.B., G.E. and G.M. acknowledge partial financial support by PRIN
2002 {\it SINTESI}.


\begin{thebibliography}{200}

\bibitem{ford} G.W. Ford, J.T. Lewis and R.F. O'Connell, Phys.
Rev. {\bf A 37}, 4419 (1988).

\bibitem{senit} I.R. Senitzky, Phys. Rev. {\bf 119}, 670 (1960).

\bibitem{peier} R. E. Peierls, Proc. Roy. Soc. (London), {\bf A 214},
143 (1952).

\bibitem{dewitt} B.S. DeWitt, {\it Dynamical Theory of Groups} {\it and
Fields}, (Gordon \& Breach, New York, 1965).

\bibitem{maro} D. Marolf, Ann. Phys. (N.Y.) {\bf 236}, 392 (1994).

\bibitem{duts} M.D\"utsch and K. Fredenhagen, The Master Ward
Identity and Generalized Schwinger--Dyson Equation in Classical
field theory, hep-th/0211242.

\bibitem{bimonte} G. Bimonte, G. Esposito, G. Marmo and C.
Stornaiolo, Peierls Brackets in Field Theory, hep-th/0301113, to
appear in Int. J. Mod. Phys. {\bf A}.

\bibitem{nelson} E. Nelson, Phys. Rev. {\bf 150}, 1079 (1966) and
references therein.

\end{thebibliography}
\end{document}